\documentclass[ a4paper,
 10pt,
 onecolumn,
 superscriptaddress,
 accepted=2010-10-02]{quantumarticle}
\pdfoutput = 1

\usepackage[utf8]{inputenc}
\usepackage[english]{babel}
\usepackage[T1]{fontenc}
\usepackage{hyperref}
\usepackage[numbers,sort]{natbib}

\usepackage{lipsum}
\usepackage{mdframed}
\usepackage{booktabs}
\usepackage{longtable}
\usepackage{enumitem}
\usepackage{xcolor}
\usepackage{graphicx}
\usepackage{amsfonts,amsmath,amssymb,amsthm}
\usepackage{array}
\usepackage{verbatim}
\usepackage{scalerel}
\usepackage{pgfplots}
\usepackage{tikz}
\hypersetup{colorlinks=false,pdfborder={0 0 0}}
\usepackage{appendix}

\newcommand{\tr}{\operatorname{tr}}
\newcommand{\sket}[1]{{\ensuremath{\lvert#1\rangle}}}
\newcommand{\lket}[1]{{\ensuremath{\left\lvert#1\right\rangle}}}
\newcommand{\ket}[1]{\if@display\lket{#1}\else\sket{#1}\fi}
\newcommand{\sbra}[1]{{\ensuremath{\langle#1\rvert}}}
\newcommand{\lbra}[1]{{\ensuremath{\left\langle#1\right\rvert}}}
\newcommand{\bra}[1]{\if@display\lbra{#1}\else\sbra{#1}\fi}
\newcommand{\sbraket}[2]{{\ensuremath{\langle#1\rvert#2\rangle}}}
\newcommand{\lbraket}[2]{{\ensuremath{\left\langle#1\!\left\rvert\vphantom{#1}#2\right.\!\right\rangle}}}
\newcommand{\braket}[2]{\if@display\lbraket{#1}{#2}\else\sbraket{#1}{#2}\fi}

\newcommand{\sketbra}[2]{{\ensuremath{\lvert #1\rangle\!\langle #2\rvert}}}
\newcommand{\lketbra}[2]{{\ensuremath{\left\lvert #1\right\rangle\!\!\left\langle #2\right\rvert}}}
\newcommand{\ketbra}[2]{\if@display\lketbra{#1}{#2}\else\sketbra{#1}{#2}\fi}

\newcommand{\proj}[1]{\ketbra{#1}{#1}}
\newcommand{\A}{\mathsf{A}}
\newcommand{\B}{\mathsf{B}}
\newcommand{\K}{\mathsf{K}}
\newcommand{\M}{\mathsf{M}}
\newcommand{\E}{\mathsf{E}}

\newcommand{\tA}{\text{A}}
\newcommand{\tB}{\text{B}}

\newcommand{\tp}{\otimes}

\makeatletter
\newtheorem*{rep@theorem}{\rep@title}
\newcommand{\newreptheorem}[2]{%
\newenvironment{rep#1}[1]{%
 \def\rep@title{#2 \ref{##1}}%
 \begin{rep@theorem}}%
 {\end{rep@theorem}}}
\makeatother

\theoremstyle{plain}
\newreptheorem{thm}{Theorem}

\newtheorem{fakt}{Fact}

\theoremstyle{definition}

\theoremstyle{remark}

\begin{document}

\title{Bounding sets of sequential quantum correlations and device-independent randomness certification}
        %Semidefinite programming hierarchies for sequential Bell scenarios and device-independent randomness certification
          %\flavio{Characterizing the set of sequential Bell correlations and applications to device-independent randomness certification}}

\author{Joseph Bowles}
\affiliation{ICFO-Institut de Ciencies Fotoniques, The Barcelona Institute of Science and Technology, 08860 Castelldefels (Barcelona), Spain}
\author{Flavio Baccari}
\affiliation{ICFO-Institut de Ciencies Fotoniques, The Barcelona Institute of Science and Technology, 08860 Castelldefels (Barcelona), Spain}
\affiliation{Max-Planck-Institut f\"ur Quantenoptik, Hans-Kopfermann-Stra{\ss}e 1, 85748 Garching, Germany}
\author{Alexia Salavrakos}
\affiliation{ICFO-Institut de Ciencies Fotoniques, The Barcelona Institute of Science and Technology, 08860 Castelldefels (Barcelona), Spain}

\begin{abstract}
An important problem in quantum information theory is that of bounding sets of correlations that arise from making local measurements on entangled states of arbitrary dimension. Currently, the best-known method to tackle this problem is the NPA hierarchy; an infinite sequence of semidefinite programs that provides increasingly tighter outer approximations to the desired set of correlations. In this work we consider a more general scenario in which one performs sequences of local measurements on an entangled state of arbitrary dimension. We show that a simple adaptation of the original NPA hierarchy provides an analogous hierarchy for this scenario, with comparable resource requirements and convergence properties. We then use the method to tackle some problems in device-independent quantum information. First, we show how one can robustly certify over 2.3 bits of device-independent local randomness from a two-quibt state using a sequence of measurements, going beyond the theoretical maximum of two bits that can be achieved with non-sequential measurements. Finally, we show tight upper bounds to two previously defined tasks in sequential Bell test scenarios. 
%We present a method to bound sets of probability distributions that arise from making sequences of local measurements on entangled quantum systems of arbitrary dimension. In the non-sequential scenario, this can be achieved via the celebrated NPA hierarchy. Here, we show that a simple adaptation of the NPA hierarchy provides a corresponding hierarchy of semidefinite programs that provides a sequence increasingly tighter outer approximation to the desired set of correlations. The method has comparable resource requirements to the NPA hierarchy and satisfies analogous convergence properties. We then use the method to tackle some problems in device-independent quantum information. First, we show how one can robustly certify over 2.3 bits of device-independent local randomness from a two-quibt state using a sequence of measurements, going beyond the theoretical maximum of 2 bits that can be achieved with non-sequential measurements. Finally, we provide upper bounds to two previously defined tasks in sequential Bell test scenarios.
\end{abstract}

\maketitle

\begin{figure}[t]
    \centering
    \includegraphics{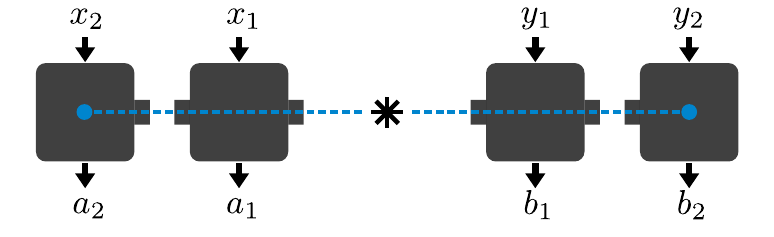}
    \caption{A sequential Bell scenario in which both parties perform a sequence of two measurements on their halves of a bipartite quantum state. In this work we develop methods to characterise the sets of probability distributions that can arise in such scenarios involving arbitrary numbers of parties in each sequence.}
    \label{fig:scenario}
\end{figure}

\section{Introduction}
The correlations between outcomes of local measurements made on entangled quantum systems are known to exhibit a rich structure. Firstly, they are generally stronger than correlations attainable via classical resources, a phenomenon known as Bell nonlocality \cite{Bell1964,Brunner2014}. Secondly, sets of quantum correlations are known to contain both smooth and flat boundaries \cite{geometry,GYNI}, and there exist correlations whose realisation requires infinite-dimensional entangled states \cite{slofstra17}, even in scenarios involving small and finite alphabet sizes. 

All of this makes the problem of characterising, and optimising over, the set of quantum correlations a highly non-trivial and potentially undecidable problem.
At the same time, being able to perform an optimisation over the entire set of quantum correlations is crucial for many areas of quantum information theory, principally in the field of device-independent quantum information, where quantum systems are treated as black-boxes and one makes no assumption on the physical dimension of the underlying state.
A major breakthrough in this direction came with the discovery of the NPA-hierarchy \cite{NPA1,NPA2}, which provides a characterisation of the set of quantum correlations via a sequence of increasing tighter outer approximations, each expressed in terms of a semi-definite program (SDP).
Consequently, the NPA hierarchy has become a vital tool for the study of device-independent protocols in the standard scenario in which they are usually considered, commonly referred to as a Bell test.
There, a bipartite state is shared between two parties, each of which makes a number of local measurements in order to generate the data that is used in the protocol. 

In recent years a number of works have also considered \emph{sequential Bell test} scenarios, in which the parties make a sequence of local measurements that obey a time-ordered causal structure \cite{Gallego2014,Silva2015,Hirsch2013,Popescu1995,gisin_hidden} (see figure \ref{fig:scenario}). Such scenarios have been shown to be relevant for Bell non-locality via for example the phenomenon of hidden nonlocality \cite{Hirsch2013,Popescu1995,gisin_hidden}. As a result, sequential measurement scenarios are known to provide an advantage in device-independent randomness certification \cite{Curchod2017} and, we expect, in many other device-independent protocols. Further to this, sequential measurement scenarios also play a role in demonstrations of contextuality \cite{pusey2014anomalous} and Leggett-Garg type tests of nonclassicality \cite{budroni2013}.

%In contrast, sets of classical correlations have comparatively simple characterisations in terms of convex polytopes \cite{Brunner2014}, whose points can always be obtained using shared classical systems of finite dimension. 

 %Importantly, each set in the hierarchy is defined by a finite number of linear and positive semi-definite constraints, and can thus be characterised by a semi-definite program (SDP) \cite{boyd,parillo} and optimised over using a suitable numerical solver. Furthermore, certified upper and lower bounds to optimisation problems can be obtained via duality theorems. Consequently, the NPA hierarchy has become a vital tool in many areas of quantum information theory, principally in the field of device-independent quantum information, where quantum systems are treated as black-boxes and one typically needs to perform an optimisation over the entire set of quantum correlations in order to prove, for example, the security of the protocol. 

It is thus very desirable to develop methods to characterise the correlations arising in sequential Bell test scenarios. In this work we show that such a characterisation is possible by augmenting the original NPA hierarcy with a finite number of additional linear constraints. This provides a sequence of outer approximations to the corresponding set of correlations that can each be defined via a suitable SDP, with analogous resource requirements and convergence properties of the NPA hierarchy. We then apply our hierarchy to several problems in quantum information. First, we investigate device-independent randomness certification. We show how to use the hierarchy to robustly certify over 2.3 bits of local randomness from a two-qubit state via a simple sequential measurement strategy, thus going beyond the theoretical maximum of two bits that is achievable in non-sequential Bell scenarios. We then show that the previously studied strategies for the simultaneous violation of two CHSH inequalities \cite{Silva2015} and the violation of the sequential Bell inequality defined in \cite{Gallego2014} are both optimal for strategies of any dimension, up to numerical precision.

We note that the recent work \cite{q_inflation} also describes a sequence of SDP relaxations for generic quantum-causal networks that can be applied to the sequential structures we consider; see the discussion for further information.

\section{Preliminaries} 

\subsection{Quantum correlations}
In a standard Bell scenario, two spatially-separated players perform measurements on their local share of a bipartite state, chosen according to some random inputs $x,y = 1,\ldots,m$, and then collect the corresponding outputs $a,b = 1,\ldots,d$.
The resulting correlations $P(a,b\vert x,y)$ are called \emph{quantum}, $P(a,b\vert x,y)\in\mathcal{Q},$ if they can be written $\tr[\rho\;\A_{a}^{x}\tp\B_{b}^{y}]$ for some bipartite quantum state $\rho$ and local measurement operators $\A_{a}^{x}$ and $\B_{b}^{y}$. Here one can take $\rho$ pure and the measurements projective without loss of generality, since any measurement on a mixed state can be realised as a projective measurement on a purification of the state \cite{stinespring_1955}. Thus,
\begin{align}\label{pab}
    &P(a,b\vert x,y)\in\mathcal{Q} \iff P(a,b\vert x,y)=\bra{\psi}\A_{a}^{x}\tp\B_{b}^{y}\ket{\psi}  
    %\\[3pt]
    %&\quad\text{for some } \ket{\psi},\A_{a}^x,\B_b^y \text{ such that } \nonumber\\[3pt]
    %&\quad\quad\langle \psi \vert \psi \rangle =1, \nonumber \\
    %&\quad\quad\sum_{a}\A_{a}^x = \openone_{\tA} \;\forall x, \quad \A_{a}^x\geq 0 \;\forall a,x, \label{norm1} \\
    %&\quad\quad\sum_{b}\B_{b}^y = \openone_{\tB} \;\forall y, \quad \B_{b}^y\geq 0 \;\forall b,y. \label{norm2}
\end{align}
with
\begin{align}\label{proj1}
    \A_a^x\A_{a'}^{x}=\A_{a}^{x}\delta_{a,a'} \;\forall x,a,a'\;, \quad
    \B_b^y\B_{b'}^{y}=\B_{b}^{y}\delta_{b,b'} \;\forall y,b,b'.
\end{align}

Since the state and measurements appearing in \eqref{pab} are potentially infinite dimensional, the problem of deciding membership in, or optimising over the set $\mathcal{Q}$ is highly non-trivial. Currently, the only general purpose technique to tackle such a problem is the NPA hierarchy \cite{NPA1,NPA2}, which we will recap shortly. 

\subsection{Sequential quantum correlations}
In this work we consider sequential measurement scenarios, where a quantum system is subjected local measurements that obey a time-ordered structure (see Fig.\ \ref{fig:scenario}). Consider first a single quantum system $\ket{\psi}$, of potentially uncountable infinite dimension, that undergoes a sequence of $n$ measurements with inputs $x_i$ and outcomes $a_i$. The first measurement outcome and its corresponding post-measurement state are described by sets of Kraus operators $\{\K_{a_1,\mu_1}^{x_1}\}$. For finite dimensional systems the (sub-normalised) post-measurement state obtained after obtaining outcome $a_1$ takes the form 
\begin{align}\label{postmeas}
    \rho_{a_1\vert x_1}=\sum_{\mu_1} \K^{x_1}_{a_1,\mu_1}\proj{\psi}\K_{a_1,
    \mu_1}^{x_1\dagger},
\end{align}
with $P(a_1\vert x_1)=\tr\rho_{a_1\vert x_1}$,  $\sum_{a_1,\mu_1}\K_{a_1,\mu_1}^\dagger\K_{a_1,\mu_1}=\openone$, and where the sum over $\mu_1$ is needed since we may have multiple Kraus operators associated to a single measurement outcome. Generally, for infinite dimensional systems one replaces the sum with an integral:
\begin{align}\label{postmeas}
    \rho_{a_1\vert x_1}=\int_{\mu_1}\text{d}\mu_1\, \K^{x_1}_{a_1,\mu_1}\proj{\psi}\K_{a_1,
    \mu_1}^{x_1\dagger},
\end{align}
where again $P(a_1\vert x_1)=\tr\rho_{a_1\vert x_1}$ and $\sum_{a_1}\int \text{d}\mu_1\K_{a_1,\mu_1}^\dagger\K_{a_1,\mu_1}=\openone$. Continuing this process for the entire sequence with inputs $\mathbf{x}=(x_1,\cdots,x_n)$ and outputs $\mathbf{a}=(a_1,\cdots,a_n)$, one finds
\begin{align}\label{seq_meas_cons}
    P(\mathbf{a}\vert \mathbf{x})=\bra{\psi}\A_{\mathbf{a}}^{\mathbf{x}}\ket{\psi},\quad\quad &\A_{\mathbf{a}}^{\mathbf{x}}=
\int\cdots\int \text{d}\mu_1\cdots\text{d}\mu_n \K_{a_1,\mu_1}^{x_1 \dagger}\K_{a_2,\mu_2}^{x_2 \dagger}\cdots \K_{a_{n},\mu_{n}}^{x_{n}\dagger}\K_{a_{n},\mu_{n}}^{x_{n}} \cdots \K_{a_2,\mu_2}^{x_2}\K_{a_1,\mu_1}^{x_1},  \nonumber\\[3pt]
&\sum_{a_i}\int \text{d}\mu_i\K_{a_i,\mu_i}^{x_i \dagger} \K_{a_i,\mu_i}^{x_i} = \openone_{\tA} \;\;\forall x_i . 
\end{align}
To ease notation we have left the time-step dependence of the Kraus operators implicit. That is, $\{\K_{a_1,\mu_1}^{x_1 \dagger}\}$ and $\{\K_{a_2,\mu_2}^{x_2 \dagger}\}$ are in general different sets of operators, which is understood from the input/output indices. We define the set of \emph{sequential quantum correlations} $\mathcal{Q}_{\text{SEQ}}$ as those that arise from performing sequential measurements locally on a bipartite quantum state $\ket{\psi}$, i.e. 
\begin{align*}
P(\mathbf{a},\mathbf{b}\vert \mathbf{x},\mathbf{y})\in\mathcal{Q}_{\text{SEQ}} \iff P(\mathbf{a},\mathbf{b}\vert \mathbf{x},\mathbf{y})=\bra{\psi}\A_{\mathbf{a}}^{\mathbf{x}}\tp\B_{\mathbf{b}}^{\textbf{y}}\ket{\psi}, 
\end{align*}
where the measurement operators $\A_{\mathbf{a}}^{\mathbf{x}}$ and $\B_{\mathbf{b}}^{\mathbf{y}}$ have the sequential structure \eqref{seq_meas_cons}. Can we define a hierarchy, analogous to the NPA hierarchy for $\mathcal{Q}$, to characterise the set $\mathcal{Q}_{\text{SEQ}}$? In this work we show how this can be achieved in an efficient manner, via a simple adaptation of the original NPA hierarchy.  %At first glance this may seem difficult since in \eqref{seq_meas_cons} we have potentially infinitely many Kraus operators appearing in the sum. Nevertheless, in the following we show that a simple adaptation to the original NPA hierarchy gives a sequence of tests that provides such a hierarchy. 

\subsection{The NPA hierarchy}
Before explaining our method, we review the NPA hierachy \cite{NPA1,NPA2}. The NPA hierarchy  provides a sequence of tests, each of which checks membership in a set $\mathcal{Q}_i\supseteq\mathcal{Q}$ such that $\mathcal{Q}_1\supseteq \mathcal{Q}_2\supseteq \cdots \supseteq\mathcal{Q}$. To see how the NPA hierarchy works, consider some state and projective measurements $\ket{\psi},\A_{a}^x,\B_b^y$ with corresponding correlations $P(a,b\vert x,y)=\bra{\psi}\A_{a}^x\B_b^y\ket{\psi}$ (where $\A_a^x$ should be understood as $\A_a^x\tp\openone_\tB$ and $\B_b^y$ as $\openone_\tA\tp \B_b^y$). Define sets $S_k$, consisting of the identity operator and all products of the operators $\A_a^x$ and $\B_{b}^y$ up to degree $k$;
\begin{align}
S_1=\{\openone\}\cup_{a,x}\{\A_a^x\}\cup_{b,y}\{\B_{b}^y\},\qquad
S_{k+1}=S_k\cup_{i,j} \{S_k^{(i)}S_1^{(j)}\}
\end{align}
where $S_k^{(i)}$ is the $i^{\text{th}}$ element of $S_k$.  %$S_2=S_1\cup_{a,a',x,x'}\{\A_a^x\A_{a'}^{x'}\}\cup_{b,b',y,y'}\{\B_{b}^y\B_{b'}^{y'}\}\cup_{a,b,x,y}\{\A_{a}^x\B_{b}^{y}\}$. %, where the set union is understood to exclude any zero elements that follow from the orthogonality conditions \eqref{proj}. 
Next, define the \emph{moment matrix of order k}, $\Gamma_k$, with elements $\Gamma_k^{(i,j)}$
\begin{align}\label{gamma}
    \Gamma_k^{(i,j)}= \bra{\psi}(S_k^{(i)})^\dagger S_k^{(j)} \ket{\psi},
\end{align}
By construction, the matrix $\Gamma_k$ has the following properties:

\begin{enumerate}[label=\roman*.]
\item $\Gamma_k$ satisfies a number of linear constraints stemming from the orthogonality properties \eqref{proj1}, the normalisation of the measurement operators, and from the commutation of Alice's and Bob's operators. For example $\bra{\psi}\A_a^x \A_{a'}^x\ket{\psi}=0$ for $a\neq a'$ and $\bra{\psi}[\A_a^x, \B_{b}^y]\ket{\psi}=0$. We can write these constraints as $\tr[\Gamma_k G_i]=0$ for some suitable fixed matrices $G_i$. 

\item $\Gamma_k$ contains some elements that correspond to observable probabilities. For example $\bra{\psi}\A_a^x \B_b^y \ket{\psi}=P(a,b\vert x,y)$. We write these constraints as $\tr[\Gamma_k F_j]=P_j$, where $F_j$ are fixed matrices and $P_j$ denotes the corresponding observed probability. Similarly, taking $S_k^{(0)}=\openone$ we have $\Gamma_k^{(0,0)}=1$  since $\tr\proj{\psi}=1$. 

\item $\Gamma_k^\dagger=\Gamma_k$ and  $\Gamma_k$ is positive semi-definite (see \cite{NPA1} for a simple proof).
\end{enumerate}

Imagine that we are given some other correlation $P(a,b,\vert x,y)$ for which we want to test membership in $\mathcal{Q}$. If $P\in\mathcal{Q}$, there exists a state and measurements leading to $P$ and a corresponding matrix $\Gamma_k$ satisfying the above conditions. We thus have a necessary condition for $P\in\mathcal{Q}$:\\[7pt]
\textbf{NPA hierarchy (level k):}
\begin{align}\label{NPA}
    \text{Find }\Gamma_k \quad\text{ such that} \quad 
    &\quad\;\Gamma_k \succcurlyeq 0,\; \Gamma_k^\dagger=\Gamma_k, \; \Gamma_k^{(0,0)}=1, \\
    &\quad\tr[\Gamma_k G_i]=0 \; \forall i, \nonumber\\
    &\quad\tr[\Gamma_k F_{i}]=P_i \; \forall i. \nonumber
\end{align}
We denote the set of correlations with a positive solution to the above problem at level $k$ as $\mathcal{Q}_k$. Since the test is a necessary condition for $P\in \mathcal{Q}$ we have $\mathcal{Q}_k \supseteq \mathcal{Q}$. As the test contains only linear and positive-semidefinite constraints, it can be cast as a SDP feasibility problem and solved efficiently (in the size of the matrix $\Gamma_k$) by a suitable solver. We thus have a sequence of SDPs, each of which provides a relaxation to the problem of deciding membership in $\mathcal{Q}$. Since $\Gamma_{k}$ is a principle sub-matrix of $\Gamma_{k+1}$, one has $\Gamma_{k+1}\succcurlyeq 0 \implies \Gamma_{k}\succcurlyeq 0$ and so $\mathcal{Q}_{k+1}\supseteq\mathcal{Q}_k$. Furthermore, one can perform optimization of linear combinations of the probabilities $P_j$ over $\mathcal{Q}_k$ by removing the final constraint in \eqref{NPA} and defining a linear combination of the elements $\tr[\Gamma_k F_j]$ as an objective function of the SDP. One can then obtain certified upper and lower bounds to the problem via duality theorems of convex optimisation. In practice, relevant problems can be tackled in this way at low levels of the hierarchy that are tractable on a desktop computer. 

In principle, one can use other sets than $S_{k}$ to generate the moment matrix \eqref{gamma}, with each choice giving a different relaxation to $\mathcal{Q}$. Often, and in the examples we present later, we will use a level that is mid-way between level 1 and level 2, often called level 1+AB. This level is defined by the set
\begin{align}
    S_{1+AB}=S_1\cup_{a,x,b,y}\{\A_{a}^x\B_{b}^y\}.
\end{align}
This set defines the lowest level in the hierarchy of \cite{MoroderNPA}, and defines the so-called set of `almost quantum' correlations \cite{almostQ}. As we will see, this set is often sufficient to get non-trival and even tight bounds to relevant optimisation problems.

\section{NPA hierarchy for sequential correlations}

First, let us state our main technical result regarding the characterisation of sequential quantum correlations.

\begin{fakt}\label{eq:fact1}
A given set of correlations $P(\mathbf{a},\mathbf{b}\vert \mathbf{x},\mathbf{y})$ belongs to $\mathcal{Q}_{\text{SEQ}}$ if and only if it can be realised as $P(\mathbf{a},\mathbf{b}\vert \mathbf{x},\mathbf{y}) = \bra{\psi}\A_{\mathbf{a}}^{\mathbf{x}}\tp\B_{\mathbf{b}}^{\textbf{y}}\ket{\psi}$, with the measurement operators being projective and satisfying one-way `no-signalling' and orthogonality conditions. That is
\begin{align}\label{orthog}
\quad\A_{\mathbf{a}}^\mathbf{x}\A_{\mathbf{a'}}^\mathbf{x}=\delta_{\mathbf{a},\mathbf{a'}}\A_{\mathbf{a}}^\mathbf{x} \quad\forall \textbf{x}, \textbf{a},\textbf{a}' 
\end{align}
\begin{align}
\sum_{a_{k+1},\cdots,a_n} \A_{\textbf{a}}^{\textbf{x}}-\A_{\textbf{a}}^{\textbf{x}'}=0 \quad &\forall a_1,\dots,a_k \, , \label{noback}\\[-10pt]  & \forall \textbf{x},\textbf{x}' \text{s.t } x_i=x_i' \; (i\leq k) \nonumber \\ & 1\leq k \leq n-1 \nonumber
\end{align}
\begin{align}\label{orthog2}
    \A_{\mathbf{a}}^{\mathbf{x}}\A_{\mathbf{a'}}^\mathbf{x'}=0 \quad \forall \textbf{x},\textbf{x}',\textbf{a},\textbf{a}'\text{ s.t } \quad & x_i=x_i', \; (i\leq k), \\ &(a_1,\dots,a_k)\neq (a_1',\dots,a_k'). \nonumber \\ 
    & 1\leq k \leq n \nonumber
\end{align}
and similarly for $\B_{\mathbf{b}}^\mathbf{y}$.
\end{fakt}
Note that the projective condition \eqref{orthog} is in fact implied by the more general condition \eqref{orthog2} and so one can equivalently take only \eqref{noback} and \eqref{orthog2} in the above. 

\begin{figure}
    \centering
    \includegraphics[scale=1.0]{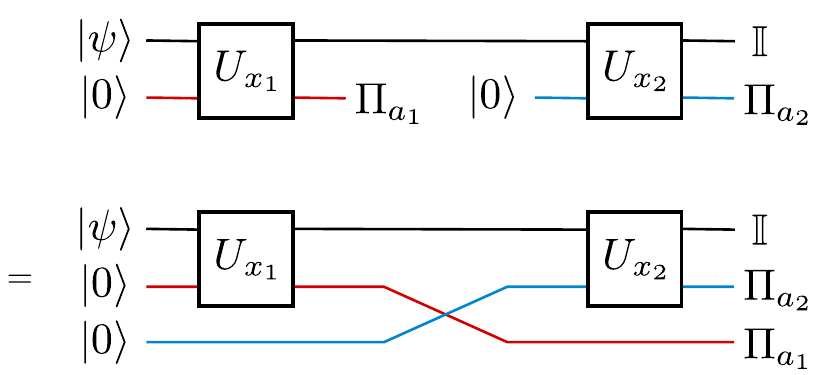}
    \caption{Top: the dilation of a sequence of two measurements. Each measurement in the sequence can be realised by appending an ancila state, performing a joint unitary operation and making a projective measurement on the ancilla space. Bottom: absorbing the ancilla states in the initial state and moving all projective measurements to the end, the scheme is equivalent to a unitary operation followed by a projective measurement. The measurement operators $\A_{\mathbf{a}}^{\mathbf{x}}$ can thus be taken projective WLOG.}
    \label{fig:main}
\end{figure}

\begin{proof}
We first prove that any correlations in $\mathcal{Q}_{\text{SEQ}}$ can be realised using measurement operators satisfying \eqref{orthog}, \eqref{noback} and \eqref{orthog2}. This can be proven by considering Stinespring dilations of the sequential measurements; see appendix \ref{app:dilation}. For example, for a sequence of two measurements (see figure \ref{fig:main}), one finds that Alice's full measurement operator can be written
\begin{align}\label{2meas}
    \A_{\mathbf{a}}^\mathbf{x}=U_1^{x_1 \dagger}U_2^{x_2 \dagger}(\openone\tp\Pi_{a_1}\tp\Pi_{a_2})U_2^{x_2}U_1^{x_1},
\end{align}
which describes a projective measurement and thus satisfies \eqref{orthog}. Here, $U_1^{x_1 \dagger}$ acts trivially (with the identity) on the third Hilbert space in the product, and $U_2^{x_2 \dagger}$ acts trivially on the second Hilbert space, as shown graphically in figure \ref{fig:main}.

The constraint \eqref{noback} is true for any set of measurement operators that are realised sequentially, as can be seen from \eqref{seq_meas_cons}. This is because it reflects the fact that the measurement operators that define the first $k$ measurements (obtained by marginalising $\A_{\mathbf{a}}^{\mathbf{x}}$ over the last $n-k$ outcomes) must be independent of the last $n-k$ inputs, since these occur later in the sequence. The Stinespring dilation of the measurement operators described previously retains the sequential structure of the measurement, and so this constraint holds for the projective measurement operators as well. For example, by summing over $a_2$ in \eqref{2meas}, one finds an operator that is independent of $x_2$. 

Finally, given the Stinespring dilations of the sequential measurements, property \eqref{orthog2} follows from the orthogonality conditions of the $\Pi_{a_j}$'s. Consider again the sequence of two measurements in \eqref{2meas}. One has for $x_1=x_1'$ and $a_1\neq a_1'$ (omitting the tensor products and the identity operator)
\begin{align}
\A_{\mathbf{a}}^{\mathbf{x}}\A_{\mathbf{a'}}^\mathbf{x'}&=U_1^{x_1\dagger}U_2^{x_2\dagger}\Pi_{a_1}\Pi_{a_2}U_2^{x_2}U_2^{x_2'\dagger}\Pi_{a_1'}\Pi_{a_2'}U_2^{x_2'}U_1^{x_1}\\
&= U_1^{x_1\dagger}U_2^{x_2\dagger}  \Pi_{a_2} U_2^{x_2}U_2^{x_2'\dagger}\Pi_{a_1}\Pi_{a_1'}\Pi_{a_2'}U_2^{x_2'}U_1^{x_1} \nonumber \\
&= 0, \nonumber
\end{align}
where we have used $[U_2^{x_2},\Pi_{a_1}]=0$. Generalising this for a general sequence we find \eqref{orthog2}

%\flavio{It now remains to show that the above constraint is sufficient to characterise precisely the set of sequential measurement operators. 
%Indeed, so far we have proven that any sequential quantum correlations can be realised with measurement operators satisfying the properties \eqref{orthog} and \eqref{noback}. However, it may still be that there exist operators satisfying the above conditions that do not have a sequential realisation in terms of product of Kraus operators. However, this turns out not to be the case.}

We now show the opposite direction, i.e.\ that any measurement operators satisfying \eqref{orthog}, \eqref{noback} and \eqref{orthog2} admit a sequential realisation. In fact, to show this we need only conditions \eqref{orthog}, \eqref{noback}. Consider a projective measurement with two input labels $x_1, x_2$ and two output labels $a_1, a_2$ defined by the measurement operators $\A_{a_1a_2}^{x_1x_2}$, and assume that the measurement operators satisfy \eqref{noback}. The measurement can be realised sequentially as follows. The first device performs a measurement with Kraus operators $\K_{a_1}^{x_1}=\sum_{a_2}\A_{a_1a_2}^{x_1x_2}$. These operators are projective and independent of $x_2$ due to \eqref{noback}. The value $x_1$ is then sent to the second device (using a classical channel) and the second device measures $\K_{a_2}^{x_2}=\sum_{a_1}\A_{a_1a_2}^{x_1x_2}$. The measurement operator describing the full sequence is therefore
\begin{align}
\K_{a_1}^{x_1\dagger}\K_{a_2}^{x_2\dagger}\K_{a_2}^{x_2}\K_{a_1}^{x_1}=\K_{a_1}^{x_1}\K_{a_2}^{x_2}\K_{a_1}^{x_1}=\sum_{a_2'}\A_{a_1a_2'}^{x_1x_2}\sum_{a_1'}\A_{a_1'a_2}^{x_1x_2}\sum_{a_2''}\A_{a_1a_2''}^{x_1x_2} = \A_{a_1a_2}^{x_1x_2} 
\end{align}
as required. In the first equality we have used the fact that the Kraus operators are Hermitian and projective by construction. The final equality follows from  $\A_{a_1a_2}^{x_1x_2}$ being projective. In the above we made use of a communication channel that we have not explicitly modelled but can be realised sequentially ; see appendix \ref{fullproof} for a proof where this channel is explicit. The general result for sequences of any length can be achieved in the same fashion by applying the same technique inductively on the sequence.

\end{proof}

% Continuing this procedure for all measurements in the sequence, we find that Alice's sequential measurement operators take the form of a unitary followed by a projective measurement (see Fig.\ ). We therefore assume without loss of generality that the operators $\A_{\mathbf{a}}^{\mathbf{x}}$ and $\B_{\mathbf{b}}^{\mathbf{y}}$ be given by sets of orthogonal projectors, i.e. 
% %
% \begin{align}\label{projcons}
%   &\A_{\mathbf{a}}^{\mathbf{x}}\A_{\mathbf{a}'}^{\mathbf{x}}=\A_{\mathbf{a}}^{\mathbf{x}}\delta_{\mathbf{a}=\mathbf{a'}} \;\forall \mathbf{x}, \nonumber\\
%       &\B_{\mathbf{b}}^{\mathbf{y}}\A_{\mathbf{b}'}^{\mathbf{y}}=\B_{\mathbf{b}}^{\mathbf{y}}\delta_{\mathbf{b}=\mathbf{b'}} \;\forall \mathbf{y}.
% \end{align}
% %

Having established Fact $1$ we may define a hierarchy of relaxations to $\mathcal{Q}_{\text{SEQ}}$ as follows.
Define moment matrices $\Gamma_k$ as in \eqref{gamma} using the projective measurement operators $\A_{\textbf{a}}^{\textbf{x}}$ and $\B_{\textbf{b}}^{\textbf{y}}$ (i.e.\ satisfying \eqref{orthog}), leading to analogous constraints to \eqref{NPA}. At this point, the relaxation is equivalent to the standard NPA hierarchy, treating the sequences of measurements as single measurements. The constraints \eqref{orthog}, \eqref{noback} and \eqref{orthog2} are linear constraints on the measurement operators and thus imply additional linear constraints on $\Gamma_k$. One can therefore add these extra constraints in the form of extra fixed matrices $G^{\text{SEQ}}_i$ to \eqref{NPA}.
This leads us to the following hierarchy for sequential quantum correlations\\

\textbf{Sequential hierarchy (level k):}
\begin{align}\label{NPAseq}
    \text{Find }\Gamma_k \quad\text{ such that} \quad 
    &\quad\;\Gamma_k \succcurlyeq 0,\; \Gamma_k^\dagger=\Gamma_k, \; \Gamma_k^{(0,0)}=1, \\
    &\quad\tr[\Gamma_k G_i]=0 \; \forall i, \nonumber\\
    &\quad\tr[\Gamma_k G^{\text{SEQ}}_i]=0 \; \forall i, \nonumber\\
    &\quad\tr[\Gamma_k F_{i}]=P_i \; \forall i. \nonumber
\end{align}
We call $\mathcal{Q}_{\text{SEQ}}^k$ the set defined at level $k$ of this hierarchy. As with the NPA hierachy, the sets $\mathcal{Q}_{\text{SEQ}}^k$ can be optimised over via SDP solvers with a comparable resource overhead. Note that due to the normalisation of measurement operators and \eqref{noback}, some of the measurement operators can be written as linear combinations of others. In practice, this means that such operators can be excluded from the sets $S_k$ (thus increasing efficiency by decreasing the size of $\Gamma_k$) since their addition will result in linear dependencies between the rows and columns of $\Gamma_k$, which do not affect the constraint $\Gamma_k\succcurlyeq 0$. This process will also introduce further constraints on the now smaller $\Gamma_k$. For example, if  $\A_{\mathbf{a}}^{\mathbf{x}}$ and $\A_{\mathbf{a}'}^{\mathbf{x}'}$ are two measurement operators that have been removed from $S_k$ through this process, then by expressing them as linear combinations of the remaining elements in $S_k$, the constraint \eqref{orthog2} gives a polynomial operator identity that implies further constraints on the moment matrix.

\subsection{Convergence of the hierarchy}
Since the conditions \eqref{noback} characterise precisely the set of sequential measurement operators and are linear constraints, one can use the same methods as in \cite{NPA2} to prove convergence of the hierarchy.
In fact, one can extract a quantum state and measurement operators from the moment matrix $\Gamma_{\infty}$ corresponding to the asymptotic level of the hierarchy. It is then straightforward to see that the added linear constraints $G^{SEQ}_i$ enforce that the extracted measurement operators satisfy property \eqref{noback}, hence having a sequential realisation.
Technically speaking, the convergence is proven to a set $\tilde{\mathcal{Q}}_\text{SEQ}\supseteq \mathcal{Q}_\text{SEQ}$. Here, $\tilde{\mathcal{Q}}_\text{SEQ}$ is the set of sequential quantum correlations where the tensor product structure is replaced by the weaker constraint that Alice and Bob's measurement operators commute, i.e. 
\begin{align}
    p(\mathbf{a},\mathbf{b}\vert \mathbf{x},\mathbf{y})\in\tilde{\mathcal{Q}}_{\text{SEQ}} \iff p(\mathbf{a},\mathbf{b}\vert \mathbf{x},\mathbf{y})=\bra{\psi}\A_{\mathbf{a}}^{\mathbf{x}}\B_{\mathbf{b}}^{\textbf{y}}\ket{\psi}.  \nonumber
\end{align}
where one has $[\A_{\mathbf{a}}^{\mathbf{x}},\B_{\mathbf{b}}^{\mathbf{y}}]=0$ for all $\textbf{a},\textbf{x},\textbf{b},\textbf{y}$ and the measurement operators have the sequential structure \eqref{seq_meas_cons}. This commuting operator formalism is used in algebraic quantum field theory \cite{AQFT}, and it is known that there exist scenarios for which $\mathcal{Q}_\text{SEQ}\subset \tilde{\mathcal{Q}}_\text{SEQ}$ \cite{slofstra2019}. 

\subsection{Relaxations of local correlations}
The hierarchy can also be used to define semidefinite programming relaxations to the set of `time ordered local correlations' defined in \cite{Gallego2014}. Such correlations are those that can be obtained by a local hidden variable model that must respect the sequential causal structure of the scenario. The idea essentially the same as that presented in \cite{Baccari17}; as we show in appendix \ref{app:local}, any hidden variable model can be seen as a special case of a quantum strategy, where all measurement operators of the same party commute. For the sequential scenario, one therefore just has to add the additional linear constraints to $\Gamma_k$ implied by the relations $[\A_{\mathbf{a}}^{\mathbf{x}},\A_{\mathbf{a'}}^{\mathbf{x'}}]=0$ and $[\B_{\mathbf{b}}^{\mathbf{y}},\B_{\mathbf{b'}}^{\mathbf{y'}}]=0$. 

\section{Applications}
In the rest of this article we use our methods to tackle a number of open questions in quantum information theory. 
Code to implement our method in python can be found in the GitLab repository \url{https://gitlab.com/josephbowles/sequentialnpa}.

\subsection{Robust device-independent certification of more that 2 bits of local randomness}
One of the most important applications of the NPA hierarchy is bounding the amount of randomness one can certify from an observed probability distribution in the device-independent setting \cite{colbeck_thesis,colbeck2011private,renner_thesis,nieto2014,nieto2018,pironio_randomness,bancal2014,acin2016optimal}. A common figure of merit that is used is the \emph{local guessing probability}, defined as the maximum probability with which an adversary---usually called Eve---could guess the value of one of the local outputs for a fixed local input. More precisely, consider the set of tripartite probability distributions $p_{\text{\tiny{ABE}}}(a,b,e\vert x,y)$ for Alice, Bob and Eve (where Eve has no input and the same output alphabet as Bob) that have a realisation in quantum theory, i.e. $p_{\text{\tiny{ABE}}}(a,b,e\vert x,y)=\bra{\psi}\A_a^x\tp \B_b^y \tp \E_e\ket{\psi}\iff p_{ABE}\in\mathcal{Q}$ for some state and measurements. Define $p_{\text{\tiny{AB}}}(a,b\vert x,y)$ and $p_{\text{\tiny{BE}}}(b,e\vert y)$ to be the corresponding marginal distributions of $p_{\text{\tiny{ABE}}}(a,b,e \vert x,y)$. The local guessing probability for Bob's input $y=y^*$ given an observed probability distribution $P_{\text{obs}}(a,b\vert x,y)$ is the best probability that Eve could guess $b$ given $y=y^*$ while simultaneously reproducing $P_\text{obs}$ when marginalising over her output. That is,
\begin{align}\label{pg}
    G(y^*)=\max_{p_{\text{\tiny{ABE}}\in\mathcal{Q}}}\;\sum_{e=1}^{\vert{b}\vert} p_{\text{\tiny{BE}}}(e,e \vert y^*) \qquad \text{such that} \qquad 
    &p_{\text{\tiny{AB}}}(a,b\vert x,y)=\sum_e p_{\text{\tiny{ABE}}}(a,b,e\vert x,y)\\&=P_{\text{obs}}(a,b\vert x,y) \nonumber
\end{align}
where $\vert{b}\vert$ is the size of Bob's output alphabet. To define the local guessing probability in the sequential scenario one imposes that the distribution $p_{\text{\tiny{ABE}}}$ be realised by a sequential quantum strategy. That is, the local guessing probability for Bob's input $\mathbf{y}^*$ given an observed distribution $P_{\text{obs}}(\mathbf{a},\mathbf{b}\vert \mathbf{x},\mathbf{y})$ becomes
\begin{align}\label{pg_seq}
    G(\mathbf{y}^*)=\max_{p_{\tiny{\text{ABE}}}}\;\sum_\mathbf{e} p_{\text{\tiny{BE}}}(\mathbf{e},\mathbf{e} \vert \mathbf{y}^*)\qquad\text{such that}\qquad
    p_{\text{\tiny{AB}}}(\mathbf{a},\mathbf{b}\vert \mathbf{x},\mathbf{y})&=\sum_{\mathbf{e}} p_{\text{\tiny{ABE}}}(\mathbf{a},\mathbf{b},\mathbf{e}\vert \mathbf{x},\mathbf{y})\\ &=P_{\text{obs}}(\mathbf{a},\mathbf{b}\vert \mathbf{x},\mathbf{y}), \nonumber
\end{align}
where the alphabet of $\mathbf{e}$ is the same as $\mathbf{b}$ and where $p_{\text{\tiny{ABE}}}$ has a sequential realisation, i.e.\
\begin{align}
    p_{\text{\tiny{ABE}}}(\mathbf{a},\mathbf{b},\mathbf{e} \vert \mathbf{x},\mathbf{y})=\bra{\psi}\A_{\mathbf{a}}^{\mathbf{x}}\otimes\B_{\mathbf{b}}^{\mathbf{y}}\otimes \E_\mathbf{e}\ket{\psi},
\end{align}
where the measurement operators $\A_{\mathbf{a}}^{\mathbf{x}}$ and $\B_{\mathbf{b}}^{\mathbf{y}}$ have the structure \eqref{seq_meas_cons}. In appendix \ref{app:random} we show how upper bounds to \eqref{pg_seq} can be obtained efficiently using our hierarchy. 

In the standard Bell scenario, the local guessing probability \eqref{pg} is always lower bounded by $1/d^2$, where $d$ is the local Hilbert space dimension of the state used to obtain the observed correlations. This follows from the fact that extremal measurements acting on a Hilbert space of dimension $d$ have at most $d^2$ outcomes \cite{extremal,acin2016optimal}. Hence, the amount of randomness, expressed as the min entropy $-\log_2 (G)$ is always lower than $2\log_2(d)$ bits. However, if one imposes the sequential structure on the local measurement one can no longer bound the number of outcomes of extremal measurements. In \cite{Curchod2017} Curchod et.\ al.\ use this to construct a protocol to obtain arbitrarily small local guessing probabilities from any two-qubit entangled pure state using a single Alice and a sequence of Bobs.  

The construction in \cite{Curchod2017} has two disadvantages however. Firstly, the number of measurements that Alice makes grows quickly with the amount of certified randomness. For example, to certify more that two bits of local randomness one needs at least 14 measurements for Alice. Secondly, although the authors prove that the protocol is noise resistant in principle, precise upper bounds on the guessing probability could not be proven for any nonzero level of noise, and the method can therefore not be used in practice. In the following we show that one can use our hierarchy to certify more than two bits of local randomness in a simple sequential scenario using only two measurements for Alice. Moreover, we use our hierarchy to calculate upper bounds to the guessing probabilities in the presence of noise, thus making the scheme experimentally relevant.

To generate the observed correlations $P_{obs}$ we consider a scenario involving one Alice and a sequence of two Bobs (that we call Bob\textsubscript{1} and Bob\textsubscript{2}), where Alice and Bob\textsubscript{1} share the two-qubit isotropic state with noise parameter $\eta$:
\begin{align}
    \rho(\eta)=(1-\eta)\proj{\phi^{\text{+}}}+\eta\,\openone/4
\end{align}
with $\ket{\phi^\text{+}}=[\ket{00}+\ket{11}]/\sqrt{2}$. Alice performs one of two measurements given by the observables $\cos\mu\,\sigma_{\text{z}}\pm\sin\mu\,\sigma_{\text{x}}$, where $\tan\mu=\sin2\epsilon$ and $\epsilon$ is a free parameter. Bob\textsubscript{1} performs one of two measurements. For $y_1=0$ he performs a projective measurement of $\sigma_\text{z}$ with Kraus operators $\proj{0}$ and $\proj{1}$. For $y_1=1$ he performs the two outcome measurement defined by the Kraus operators
\begin{align}\label{kraus}
    \K_{+}=\cos\epsilon \proj{+}+\sin\epsilon\proj{-}, \qquad
    \K_{-}=-\cos\epsilon \proj{-}+\sin\epsilon\proj{+}.
\end{align}
The parameter $\epsilon$ controls the strength of the measurement: for $\epsilon=0$, the measurement is a projective measurement in the $x$ direction; for $\epsilon=\pi/4$ the measurement is non-interacting. Bob\textsubscript{2} performs one of three measurements. For $y_2=0,1$ he performs a projective measurement of $\sigma_{\text{z}}$ or $\sigma_{\text{x}}$. For $y_2=2$ he performs the symmetric 3-outcome POVM given by the measurement operators 
\begin{align}
    \M_{b_2}=\frac{2}{3}\frac{\openone+\mathbf{v}_{b_2}\cdot\vec{\sigma}}{2}\quad b_2=0,1,2,
\end{align}
where $\mathbf{v}_{b_2}=(\sin( \frac{2\pi}{3}b_2),0,\cos(\frac{2\pi}{3}b_2))$. The inspiration for these measurements is the following. For $y_1=1$, the post measurement state shared between Alice and Bob$_2$ will be one of two partially entangled states, depending on the value of $b_1$. The correlations obtained by performing the measurements for $x,y_2=0,1$ on these states are known to self-test both of the corresponding state and measurements \cite{bamps2015sum}. We expect (although we have not proven) that this implies that the state shared between Alice and Bob$_1$ is $\ket{\phi^+}$ and the measurement for Bob$_1$ \eqref{kraus}, which essentially implies that one must have $p(b_2\vert y_2=2)=\frac{1}{3}$, leading to more than two bits of randomness. 

\begin{figure}
\centering
\pgfplotsset{scaled y ticks=false}
\begin{tikzpicture}[scale=0.96]
\begin{axis}[ 
xlabel={noise parameter $\eta$}, 
ylabel={min entropy (bits)},
xtick={0,0.02,0.04,0.06,0.08,0.10,0.12,0.14},
xticklabels={0,0.02,0.04,0.06,0.08,0.10,0.12,0.14},
%ytick pos=left,
%ylabel style={yshift=-0.5cm,xshift=0},
%xmax=2.9,ymax=2.9,xmin=0,ymin=1.3,
%legend style={at={(0.03,0.03)},anchor=south west}
] 
% \pgfplotstableread{pguess_morebits_5pi_32.dat}\loadedtable;
% \addplot+[smooth,mark options={scale=0.5,mark=*}] table {\loadedtable};

% \pgfplotstableread{pguess_morebits_6pi_32.dat}\loadedtable;
% \addplot+[smooth,mark options={scale=0.5,mark=*}] table {\loadedtable};

\pgfplotstableread{pguess_morebits_7pi_32.dat}\loadedtable;
\addplot+[smooth,mark options={scale=0.5,mark=none},line width = 1.2,skip coords between index={130}{316}] table {\loadedtable};
\addlegendentry{seq.\ strategy (level $\mathcal{Q}_{\text{SEQ}}^{\text{1+AB}}$)}

\pgfplotstableread{pguess_chsh.dat}\loadedtable;
\addplot+[smooth,mark options={scale=0.5,mark=none},line width = 1.2pt,skip coords between index={130}{316}] table {\loadedtable};
\addlegendentry{CHSH strategy (NPA level 4)}

% \legend{seq.\ strategy \newline ($\mathcal{Q}_{\text{SEQ}}$ level 1+AB), CHSH strategy (NPA level 4)}
\end{axis}
\end{tikzpicture}
\caption{\label{fig_rand} Blue: lower bound to the local randomness as a function of the noise parameter $\eta$ for our sequential measurement strategy, obtained at level 1+AB of our sequential hierarchy. Red: corresponding local randomness obtainable with the same state in a non-sequential scenario using measurements that lead to the maximal violation of the CHSH Bell inequality, obtained at level 4 of the NPA hierarchy.}
\end{figure}
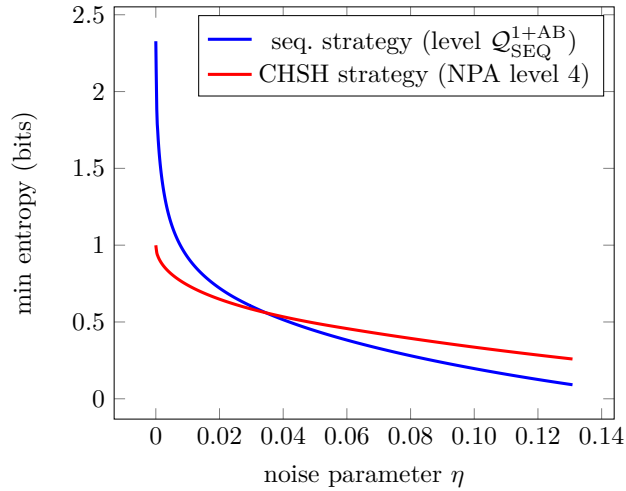

In figure \ref{fig_rand} we present upper bounds to $G(\mathbf{y}^*=(1,2))$ obtained in this way as a function of $\eta$, with $\epsilon=7\pi/32$ and calculated using level $1+AB$ of the hierarchy. For low noise, one can surpass two bits of randomness. Moreover, for close to $4\%$ noise (well within experimental reach) our strategy outperforms the non-sequential strategy where one performs the measurement that maximally violate the CHSH Bell inequality on the same state. We leave a more detailed analysis of noise including detector inefficiencies to future work. 

\subsection{Monogamy of nonlocality in sequential measurement scenarios}
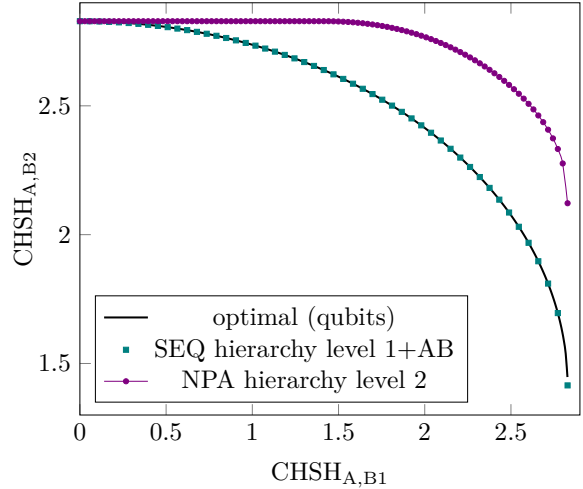
\begin{figure}
\centering
\pgfplotsset{scaled y ticks=false}
\begin{tikzpicture}[scale=0.96]
\begin{axis}[ 
xlabel={CHSH\textsubscript{A,B1}}, 
ylabel={CHSH\textsubscript{A,B2}},
ytick pos=left,
ylabel style={yshift=-0.5cm,xshift=0},
xmax=2.9,ymax=2.9,xmin=0,ymin=1.3,
legend style={at={(0.03,0.03)},anchor=south west}
] 
% \pgfplotstableread{tradeoff_level1_plot.dat}\loadedtable;
\addplot[smooth,samples=1000,thick,color=black] {sqrt(2)*(1+sqrt(1-x^2/8))};
%\addplot+[only marks,mark options={scale=0.5,mark=*},color=purple] table {\loadedtable};
§t
\pgfplotstableread{CHSH_tradeoff_level1AB_sparse2.dat}\loadedtable;
\addplot+[only marks,mark options={scale=0.5},color=teal] table {\loadedtable}; 
\legend{optimal (qubits), level 1+AB}

\pgfplotstableread{tradeoff_level2_NPA.dat}\loadedtable;
\addplot+[smooth,mark options={scale=0.5, mark=*},color=violet] table {\loadedtable}; 
\legend{optimal (qubits),  SEQ hierarchy level 1+AB, NPA hierarchy level 2}
\end{axis}

%
% \pgfplotsset{every tick label/.append style={font=\tiny}}
% \begin{axis}[
%   axis y line*=right,
%   axis x line=none,
%   ymin=0, 
%   ymax=0.05,
%   xmin=0,
%   xmax=2.9,
%   ytick={  0, 0.0025,  0.005, 0.0075},
%   yticklabels={  0, 0.0025,  0.005 , 0.0075},
%   xtick={},
%   xticklabels={},
%   color=black,
  %ylabel=error,
  %ylabel style={yshift=0cm,xshift=-2.3cm},
% ]
% \pgfplotstableread{error_data_L1AB.dat}\loadedtable
% \addplot+[mark options={scale=0.2},gray] table {\loadedtable}; 
% \end{axis}
\end{tikzpicture}
\caption{\label{tradeoff} Upper bounds on the maximum value of \text{CHSH\textsubscript{A,B2}} as a function of the value of \text{CHSH\textsubscript{A,B1}}. The values obtained at level 1+AB of the sequential hierarchy match the optimal values for qubit strategies found in \cite{Silva2015}. To show the effect of our new constraints, we plot the same bounds obtained via the standard NPA hierarchy at level 2, treating the two Bobs as a single party.}
\end{figure}

Consider a scenario involving one Alice and two Bobs, where each party has two inputs and two outputs, with inputs and outputs labelled by 0,1. The value of the CHSH Bell functional between Alice and Bob\textsubscript{1} is 
\begin{align}
    \text{CHSH\textsubscript{AB1}}=\sum_{x,y_1}\sum_{a,b_1}(-1)^{a+b_1+x\cdot y_1}P_{\text{\tiny{AB1}}}(a,b_1\vert x,y_1)
\end{align}
where $P_{\text{\tiny{AB1}}}$ is the marginal distribution between Alice and Bob\textsubscript{1}. We may define the average CHSH Bell functional between Alice and Bob\textsubscript{2} as
\begin{align}\nonumber
    \text{CHSH\textsubscript{AB2}}=\frac{1}{2}\sum_{b_1,y_1}\sum_{x,y_2}\sum_{a,b_2}(-1)^{a+b_2+x\cdot y_2}P(a,b_1,b_2\vert x,y_1,y_2),
\end{align}
i.e.\ the CHSH Bell functional between Alice and Bob\textsubscript{2}, averaged over $b_1$ and a uniform choice of $y_1$. The values of CHSH\textsubscript{AB1} and CHSH\textsubscript{AB2} are subject to monogamy due to both the monogamy of correlations and the sequential measurement constraints. Silva et.\ al.\ investigate this in \cite{Silva2015}, finding that for two-qubit systems, the optimal trade-off satisfies
\begin{align}\label{tradeoffeq}
    \text{CHSH\textsubscript{AB2}}\leq \sqrt{2}\left(1+\sqrt{1-\frac{(\text{CHSH\textsubscript{AB1}})^2}{8}}\right),
\end{align}
which can be saturated with an appropriate choice of measurements. We use the sequential NPA hierarchy to investigate this trade-off for systems of general dimension. We numerically maximise the value of CHSH\textsubscript{AB2} conditioned on values of CHSH\textsubscript{AB1} at level 1+AB of the hierarchy (see figure \ref{tradeoff}). We find that the values obtained match those of \eqref{tradeoffeq} up to the precision of the SDP solver. Thus, we conjecture that the strategies presented in \cite{Silva2015} are optimal for any dimension. This is somewhat surprising since one may expect to gain an advantage from higher dimensional systems. For example, it would allow Bob\textsubscript{1} to communicate perfectly the value of $y_1$ and $b_1$ to Bob\textsubscript{2}, which in principle could increase the value of CHSH\textsubscript{A,B2}. 

\subsection{Tight bounds on sequential Bell inequalities}
In \cite{Gallego2014} Gallego et.\ al.\ present a Bell inequality (see equation 51 therein) that defines a facet of the set of correlations that admit a sequential time-ordered local model. The scenario involves one Alice and two Bobs, with each party performing one of two dichotomic measurements. The Bell inequality is constructed as follows. Define the correlators
\begin{align}
    &\langle A_x B^2_{y_1y_2}\rangle = P(a\cdot b_2 = +1 \vert x,y_1,y_2) - P(a\cdot b_2 = -1 \vert x,y_1,y_2) \nonumber\\
    &\langle A_x B^1_{y_1}B^2_{y_1y_2}\rangle = P(a\cdot b_1 \cdot b_2 = +1 \vert x,y_1,y_2) - P(a\cdot b_1 \cdot b_2 = -1 \vert x,y_1,y_2).
\end{align}
The inequality is given by
\begin{align}
    \mathcal{I}=\langle A_0(B-B')-A_1(B+B')\rangle \leq 2
\end{align}
where 
\begin{align}
    &B=\frac{1}{2}[(1+B_0^1)B_{01}^2 - (1-B_0^1)B_{00}^2] \nonumber \\
    &B'=\frac{1}{2}[(1-B_1^1)B_{11}^2 + (1+B_{1}^1)B_{10}^2]
\end{align}
and the bound 2 holds for sequential time-ordered local correlations.

The authors show that it is possible to violate the inequality up to a value of $2\sqrt{2}$ using a sequential quantum strategy, providing a lower bound to the maximum violation using a sequential quantum strategy. Using our hierarchy at level 1+AB, we are able to certify a corresponding upper bound that agrees with the value $2\sqrt{2}$ up to the precision of the SDP solver. We therefore expect that the strategy given in \cite{Gallego2014} is optimal for this inequality.

\section{Discussion}
We have presented a general method to bound sets of correlations arising from performing sequential measurements on entangled quantum states. 
Our techniques can be seen as part of a collection of works that extend the original applicability of the NPA hierarchy to scenarios of restricted dimension \cite{NPAdimension1,NPAdimension2} and entanglement \cite{MoroderNPA}, classicality \cite{Baccari17}, and modified causality \cite{pozasNPA,q_inflation}.

We note that the techniques described in \cite{q_inflation} can in principle deal with the sequential causal structures considered in this work. More specifically, one could use their method to treat `quantum exogenous' variables by explicitly using the unitaries in \eqref{2meas} as operators in the generating set $S_k$ and defining a resulting relaxation. This method is significantly less efficient however since one needs to go to high levels (with large moment matrices) of the corresponding relaxation, and no convergence properties are proven. Given these points, it would thus be interesting to study whether our method could be extended to other causal scenarios, or be used to improve the efficiency of the method in \cite{q_inflation}. For example, can our method be applied to give a convergent hierarchy for a generic  causal structure involving latent quantum variables?

The NPA hierarchy is often used as a numerical method to bound fidelities in self-testing protocols \cite{STreview}. One avenue of research would therefore be to investigate whether sequential measurement scenarios can improve self-testing fidelity bounds, by adapting the current method to our hierarchy, or to investigate the self-testing of quantum channels, to which sequential measurement scenarios are naturally related. Finally, it would also be interesting to use our method to investigate to what extent sequential measurements can improve other device-independent protocols. For example, can our advantages in local guessing probability be translated to practical improvements to rates in randomness extraction or quantum key distribution protocols? 

\section*{Acknowledgements} 
We thank Erik Woodhead for pointing out equation \eqref{orthog2} and Flavien Hirsch for inspiring preliminary discussions. We also thank Daniel Cavalcanti, Florian Curchod, Dr.\ Bibounde,  Antonio Acin,  Remigiusz Augusiak, Marco Tulio Quintino and Peter Wittek for discussions throughout the project.

All authors acknowledge funding from the Spanish MINECO (QIBEQI FIS2016-80773-P, Severo Ochoa SEV-2015-0522, a Severo Ochoa PhD fellowship), Fundacio Cellex, Generalitat de Catalunya (SGR 1381 and CERCA Programme). JB acknowldges funding from the AXA Chair in Quantum Information Science, Juan de la Cierva-formación and the EU Quantum Flagship project QRANGE. FB acknowledges the support from the Deutsche Forschungsgemeinschaft (DFG, German Research Foundation) - Project number 414325145 in the framework of the Austrian Science Fund (FWF): SFB F71.

\bibliographystyle{unsrtnat}
\bibliography{bibliography}

\begin{thebibliography}{35}
\providecommand{\natexlab}[1]{#1}
\providecommand{\url}[1]{\texttt{#1}}
\expandafter\ifx\csname urlstyle\endcsname\relax
  \providecommand{\doi}[1]{doi: #1}\else
  \providecommand{\doi}{doi: \begingroup \urlstyle{rm}\Url}\fi

\bibitem[Bell(1964)]{Bell1964}
J.~S. Bell.
\newblock On the {E}instein-{P}odolsky-{R}osen paradox.
\newblock \emph{Physics}, 1:\penalty0 195, 1964.

\bibitem[Brunner et~al.(2014)Brunner, Cavalcanti, Pironio, Scarani, and
  Wehner]{Brunner2014}
N.~Brunner, D.~Cavalcanti, S.~Pironio, V.~Scarani, and S.~Wehner.
\newblock Bell nonlocality.
\newblock \emph{Rev. Mod. Phys.}, 86:\penalty0 839--840, 2014.
\newblock \doi{10.1103/RevModPhys.86.839}.

\bibitem[Goh et~al.(2018)Goh, Kaniewski, Wolfe, V\'ertesi, Wu, Cai, Liang, and
  Scarani]{geometry}
Koon~Tong Goh, J\ifmmode \mbox{\k{e}}\else~\k{e}drzej Kaniewski, Elie Wolfe,
  Tam\'as V\'ertesi, Xingyao Wu, Yu~Cai, Yeong-Cherng Liang, and Valerio
  Scarani.
\newblock Geometry of the set of quantum correlations.
\newblock \emph{Phys. Rev. A}, 97:\penalty0 022104, Feb 2018.
\newblock \doi{10.1103/PhysRevA.97.022104}.

\bibitem[Almeida et~al.(2010)Almeida, Bancal, Brunner, Ac\'{\i}n, Gisin, and
  Pironio]{GYNI}
Mafalda~L. Almeida, Jean-Daniel Bancal, Nicolas Brunner, Antonio Ac\'{\i}n,
  Nicolas Gisin, and Stefano Pironio.
\newblock Guess your neighbor's input: A multipartite nonlocal game with no
  quantum advantage.
\newblock \emph{Phys. Rev. Lett.}, 104:\penalty0 230404, Jun 2010.
\newblock \doi{10.1103/PhysRevLett.104.230404}.

\bibitem[Slofstra(2017)]{slofstra17}
William Slofstra.
\newblock {T}he set of quantum correlations is not closed, 2017.
\newblock URL \url{https://arxiv.org/abs/1703.08618}.
\newblock arXiv:1703.08618v2.

\bibitem[Navascu\'es et~al.(2007)Navascu\'es, Pironio, and Ac\'{\i}n]{NPA1}
M.~Navascu\'es, S.~Pironio, and A.~Ac\'{\i}n.
\newblock Bounding the set of quantum correlations.
\newblock \emph{Phys. Rev. Lett.}, 98:\penalty0 010401, 2007.
\newblock \doi{10.1103/PhysRevLett.98.010401}.

\bibitem[Navascu\'es et~al.(2008)Navascu\'es, Pironio, and Ac\'in]{NPA2}
M.~Navascu\'es, S.~Pironio, and A.~Ac\'in.
\newblock A convergent hierarchy of semidefinite programs characterizing the
  set of quantum correlations.
\newblock \emph{New Journal of Physics}, 10\penalty0 (7):\penalty0 073013,
  2008.
\newblock \doi{10.1088/1367-2630/10/7/073013}.

\bibitem[Gallego et~al.(2014)Gallego, Würflinger, Chaves, Ac{\'{\i}}n, and
  Navascu{\'{e}}s]{Gallego2014}
Rodrigo Gallego, Lars~Erik Würflinger, Rafael Chaves, Antonio Ac{\'{\i}}n, and
  Miguel Navascu{\'{e}}s.
\newblock Nonlocality in sequential correlation scenarios.
\newblock \emph{New Journal of Physics}, 16\penalty0 (3):\penalty0 033037, mar
  2014.
\newblock \doi{10.1088/1367-2630/16/3/033037}.

\bibitem[Silva et~al.(2015)Silva, Gisin, Guryanova, and Popescu]{Silva2015}
Ralph Silva, Nicolas Gisin, Yelena Guryanova, and Sandu Popescu.
\newblock Multiple observers can share the nonlocality of half of an entangled
  pair by using optimal weak measurements.
\newblock \emph{Phys. Rev. Lett.}, 114:\penalty0 250401, Jun 2015.
\newblock \doi{10.1103/PhysRevLett.114.250401}.

\bibitem[Hirsch et~al.(2013)Hirsch, Quintino, Bowles, and Brunner]{Hirsch2013}
Flavien Hirsch, Marco~T\'ulio Quintino, Joseph Bowles, and Nicolas Brunner.
\newblock Genuine hidden quantum nonlocality.
\newblock \emph{Phys. Rev. Lett.}, 111:\penalty0 160402, Oct 2013.
\newblock \doi{10.1103/PhysRevLett.111.160402}.

\bibitem[Popescu(1995)]{Popescu1995}
Sandu Popescu.
\newblock Bell's inequalities and density matrices: Revealing ``hidden''
  nonlocality.
\newblock \emph{Phys. Rev. Lett.}, 74:\penalty0 2619--2622, Apr 1995.
\newblock \doi{10.1103/PhysRevLett.74.2619}.

\bibitem[Gisin(1996)]{gisin_hidden}
N.~Gisin.
\newblock Hidden quantum nonlocality revealed by local filters.
\newblock \emph{Physics Letters A}, 210\penalty0 (3):\penalty0 151 -- 156,
  1996.
\newblock ISSN 0375-9601.
\newblock \doi{https://doi.org/10.1016/S0375-9601(96)80001-6}.

\bibitem[Curchod et~al.(2017)Curchod, Johansson, Augusiak, Hoban, Wittek, and
  Ac\'{\i}n]{Curchod2017}
F.~J. Curchod, M.~Johansson, R.~Augusiak, M.~J. Hoban, P.~Wittek, and
  A.~Ac\'{\i}n.
\newblock Unbounded randomness certification using sequences of measurements.
\newblock \emph{Phys. Rev. A}, 95:\penalty0 020102, Feb 2017.
\newblock \doi{10.1103/PhysRevA.95.020102}.

\bibitem[Pusey(2014)]{pusey2014anomalous}
Matthew~F Pusey.
\newblock Anomalous weak values are proofs of contextuality.
\newblock \emph{Physical review letters}, 113\penalty0 (20):\penalty0 200401,
  2014.
\newblock \doi{10.1103/PhysRevLett.113.200401}.

\bibitem[Budroni et~al.(2013)Budroni, Moroder, Kleinmann, and
  G{\"u}hne]{budroni2013}
Costantino Budroni, Tobias Moroder, Matthias Kleinmann, and Otfried G{\"u}hne.
\newblock Bounding temporal quantum correlations.
\newblock \emph{Physical review letters}, 111\penalty0 (2):\penalty0 020403,
  2013.
\newblock \doi{10.1103/PhysRevLett.111.020403}.

\bibitem[Wolfe et~al.(2019)Wolfe, Pozas-Kerstjens, Grinberg, Rosset, Ac{\'\i}n,
  and Navascues]{q_inflation}
Elie Wolfe, Alejandro Pozas-Kerstjens, Matan Grinberg, Denis Rosset, Antonio
  Ac{\'\i}n, and Miguel Navascues.
\newblock Quantum inflation: A general approach to quantum causal
  compatibility.
\newblock \emph{arXiv preprint arXiv:1909.10519}, 2019.
\newblock URL \url{https://arxiv.org/abs/1909.10519}.

\bibitem[Stinespring(1955)]{stinespring_1955}
W.~Stinespring.
\newblock Positive functions on ${C}\sp *$-algebras.
\newblock \emph{Proceedings of the American Mathematical Society}, 6\penalty0
  (2):\penalty0 211–211, Jan 1955.
\newblock \doi{10.1090/s0002-9939-1955-0069403-4}.

\bibitem[Moroder et~al.(2013)Moroder, Bancal, Liang, Hofmann, and
  G\"uhne]{MoroderNPA}
Tobias Moroder, Jean-Daniel Bancal, Yeong-Cherng Liang, Martin Hofmann, and
  Otfried G\"uhne.
\newblock Device-independent entanglement quantification and related
  applications.
\newblock \emph{Phys. Rev. Lett.}, 111:\penalty0 030501, Jul 2013.
\newblock \doi{10.1103/PhysRevLett.111.030501}.

\bibitem[Navascu{\'e}s et~al.(2015)Navascu{\'e}s, Guryanova, Hoban, and
  Ac{\'\i}n]{almostQ}
Miguel Navascu{\'e}s, Yelena Guryanova, Matty~J Hoban, and Antonio Ac{\'\i}n.
\newblock Almost quantum correlations.
\newblock \emph{Nature communications}, 6\penalty0 (1):\penalty0 1--7, 2015.
\newblock \doi{10.1038/ncomms7288}.

\bibitem[Haag and Kastler(1964)]{AQFT}
Rudolf Haag and Daniel Kastler.
\newblock An algebraic approach to quantum field theory.
\newblock \emph{Journal of Mathematical Physics}, 5\penalty0 (7):\penalty0
  848--861, 1964.
\newblock \doi{10.1063/1.1704187}.

\bibitem[Slofstra(2020)]{slofstra2019}
William Slofstra.
\newblock Tsirelson’s problem and an embedding theorem for groups arising
  from non-local games.
\newblock \emph{Journal of the American Mathematical Society 33}, 2020.
\newblock \doi{10.1090/jams/929}.

\bibitem[Baccari et~al.(2017)Baccari, Cavalcanti, Wittek, and
  Ac\'{\i}n]{Baccari17}
F.~Baccari, D.~Cavalcanti, P.~Wittek, and A.~Ac\'{\i}n.
\newblock Efficient device-independent entanglement detection for multipartite
  systems.
\newblock \emph{Phys. Rev. X}, 7:\penalty0 021042, Jun 2017.
\newblock \doi{10.1103/PhysRevX.7.021042}.

\bibitem[Colbeck(2009)]{colbeck_thesis}
Roger Colbeck.
\newblock Quantum and relativistic protocols for secure multi-party
  computation.
\newblock \emph{arXiv preprint arXiv:0911.3814}, 2009.
\newblock URL \url{https://arxiv.org/abs/0911.3814}.

\bibitem[Colbeck and Kent(2011)]{colbeck2011private}
Roger Colbeck and Adrian Kent.
\newblock Private randomness expansion with untrusted devices.
\newblock \emph{Journal of Physics A: Mathematical and Theoretical},
  44\penalty0 (9):\penalty0 095305, 2011.
\newblock \doi{10.1088/1751-8113/44/9/095305}.

\bibitem[Renner(2008)]{renner_thesis}
Renato Renner.
\newblock Security of quantum key distribution.
\newblock \emph{International Journal of Quantum Information}, 6\penalty0
  (01):\penalty0 1--127, 2008.
\newblock \doi{10.1142/S0219749908003256}.

\bibitem[Nieto-Silleras et~al.(2014)Nieto-Silleras, Pironio, and
  Silman]{nieto2014}
Olmo Nieto-Silleras, Stefano Pironio, and Jonathan Silman.
\newblock Using complete measurement statistics for optimal device-independent
  randomness evaluation.
\newblock \emph{New Journal of Physics}, 16\penalty0 (1):\penalty0 013035,
  2014.
\newblock \doi{10.1088/1367-2630/16/1/013035}.

\bibitem[Nieto-Silleras et~al.(2018)Nieto-Silleras, Bamps, Silman, and
  Pironio]{nieto2018}
Olmo Nieto-Silleras, C{\'e}dric Bamps, Jonathan Silman, and Stefano Pironio.
\newblock Device-independent randomness generation from several bell
  estimators.
\newblock \emph{New journal of physics}, 20\penalty0 (2):\penalty0 023049,
  2018.
\newblock \doi{10.1088/1367-2630/aaaa06}.

\bibitem[Pironio et~al.(2010)Pironio, Ac{\'\i}n, Massar, de~La~Giroday,
  Matsukevich, Maunz, Olmschenk, Hayes, Luo, Manning,
  et~al.]{pironio_randomness}
Stefano Pironio, Antonio Ac{\'\i}n, Serge Massar, A~Boyer de~La~Giroday,
  Dzmitry~N Matsukevich, Peter Maunz, Steven Olmschenk, David Hayes, Le~Luo,
  T~Andrew Manning, et~al.
\newblock Random numbers certified by bell’s theorem.
\newblock \emph{Nature}, 464\penalty0 (7291):\penalty0 1021, 2010.
\newblock \doi{10.1038/nature09008}.

\bibitem[Bancal et~al.(2014)Bancal, Sheridan, and Scarani]{bancal2014}
Jean-Daniel Bancal, Lana Sheridan, and Valerio Scarani.
\newblock More randomness from the same data.
\newblock \emph{New Journal of Physics}, 16\penalty0 (3):\penalty0 033011,
  2014.
\newblock \doi{10.1088/1367-2630/16/3/033011}.

\bibitem[Ac{\'\i}n et~al.(2016)Ac{\'\i}n, Pironio, V{\'e}rtesi, and
  Wittek]{acin2016optimal}
Antonio Ac{\'\i}n, Stefano Pironio, Tam{\'a}s V{\'e}rtesi, and Peter Wittek.
\newblock Optimal randomness certification from one entangled bit.
\newblock \emph{Physical Review A}, 93\penalty0 (4):\penalty0 040102, 2016.
\newblock \doi{10.1103/PhysRevA.93.040102}.

\bibitem[D'Ariano et~al.(2005)D'Ariano, Presti, and Perinotti]{extremal}
Giacomo~Mauro D'Ariano, Paoloplacido~Lo Presti, and Paolo Perinotti.
\newblock Classical randomness in quantum measurements.
\newblock \emph{Journal of Physics A: Mathematical and General}, 38\penalty0
  (26):\penalty0 5979, 2005.
\newblock \doi{10.1088/0305-4470/38/26/010}.

\bibitem[Bamps and Pironio(2015)]{bamps2015sum}
C{\'e}dric Bamps and Stefano Pironio.
\newblock Sum-of-squares decompositions for a family of
  clauser-horne-shimony-holt-like inequalities and their application to
  self-testing.
\newblock \emph{Physical Review A}, 91\penalty0 (5):\penalty0 052111, 2015.
\newblock \doi{10.1103/PhysRevA.91.052111}.

\bibitem[Navascu\'es and V\'ertesi(2015)]{NPAdimension1}
Miguel Navascu\'es and Tam\'as V\'ertesi.
\newblock Bounding the set of finite dimensional quantum correlations.
\newblock \emph{Phys. Rev. Lett.}, 115:\penalty0 020501, Jul 2015.
\newblock \doi{10.1103/PhysRevLett.115.020501}.

\bibitem[Navascu\'es et~al.(2014)Navascu\'es, de~la Torre, and
  V\'ertesi]{NPAdimension2}
Miguel Navascu\'es, Gonzalo de~la Torre, and Tam\'as V\'ertesi.
\newblock Characterization of quantum correlations with local dimension
  constraints and its device-independent applications.
\newblock \emph{Phys. Rev. X}, 4:\penalty0 011011, Jan 2014.
\newblock \doi{10.1103/PhysRevX.4.011011}.

\bibitem[{\v{S}}upi{\'{c}} and Bowles(2020)]{STreview}
Ivan {\v{S}}upi{\'{c}} and Joseph Bowles.
\newblock Self-testing of quantum systems: a review.
\newblock \emph{{Quantum}}, 4:\penalty0 337, September 2020.
\newblock ISSN 2521-327X.
\newblock \doi{10.22331/q-2020-09-30-337}.

\end{thebibliography}

\appendix
\section{Stinespring dilation of sequential measurement}\label{app:dilation}
Consider for simplicity a sequence of two measurements with inputs $x_1,x_2$ and outputs $a_1,a_2$ on a single quantum system $\ket{\psi}$ that is described by Kraus operators $\K_{a_j,i_j}^{x_j}$. The sequence of measurement can be realised as follows. Introduce ancilla spaces $\tA_1'$ and $\tA_1''$ and the ancilla state $\ket{0}=\ket{0}_{\tA_1'}\ket{0}_{\tA_1''}$. Define an operator $U_1^{x_1}$ via its action on the state $\ket{\psi}\ket{0}$ as
\begin{align}
    U_1^{x_1}\ket{\psi}\ket{0}=\sum_{a_1}\int_{\mu_1}\text{d}\mu_1(\K_{a_1,\mu_1}^{x_1}\ket{\psi})\ket{a_1}\ket{\mu_1}.
\end{align}
One has $\bra{\phi}\bra{0}\bra{0}U_1^{x_1 \dagger}U_1^{x_1}\ket{\psi}\ket{0}\ket{0}=\langle\phi\vert\psi\rangle$ for all $\ket{\psi},\ket{\phi}$. It follows that $U_1^{x_1}$ can be extended to a unitary operator acting on  $\ket{\psi}\ket{0}$. Measure the $\tA'_1$ space in the $\ket{a_1}$ basis, obtaining outcome $a_1$. Conditioning on outcome $a_1$ and tracing out the $\tA_1'$ and $\tA_1''$ spaces, one finds \eqref{postmeas}. We have thus reproduced the first measurement in the sequence. Introducing a fresh ancilla and repeating this for the second measurement in the sequence we find
\begin{align}
    \A_{\mathbf{a}}^\mathbf{x}=U_1^{x_1 \dagger}U_2^{x_2 \dagger}(\openone\tp\Pi_{a_1}\tp\Pi_{a_2})U_2^{x_2}U_1^{x_1},
\end{align}
where the $\Pi_{a_i}$'s are projectors onto the corresponding spaces. The full measurement $\A_{\mathbf{a}}^\mathbf{x}$ is thus projective. We may repeat this process for a sequence of arbitrary length, and hence $\A_{\mathbf{a}}^\mathbf{x}$ can be taken to be projective without loss of generality.

\section{Detailed proof of fact 1}\label{fullproof}
Here we give a proof of the reverse direction of fact 1, where we explicitly model the communication channel in the Kraus operators. Enlarge the system via an ancilla state so that the full state is $\ket{\psi}\tp\ket{0}$. This space will be used as a communication channel in the following. The first device performs a measurement with Kraus operators $\K_{a_1}^{x_1}=\sum_{a_2}\A_{a_1a_2}^{x_1x_2}\tp V_{x_1}$, where $V_{x_1}$ is a unitary operator that maps $\ket{0}$ to $\ket{x_1}$ (for example, if $x_1=0,1$ then $V_{x_1}=\sigma_x^{x_1}$). These operators are independent of $x_2$ due to \eqref{noback}. The second device measures (projective) Kraus operators $\K_{a_2}^{x_2}=\sum_{x_1,a_1}\A_{a_1a_2}^{x_1x_2}\tp\proj{x_1}$. The measurement operator describing the full sequence is therefore
\begin{align}
\K_{a_1}^{x_1\dagger}\K_{a_2}^{x_2\dagger}\K_{a_2}^{x_2}\K_{a_1}^{x_1}=&(\sum_{a_2'}\A_{a_1a_2'}^{x_1x_2}\tp V_{x_1}^\dagger)(\sum_{x_1',a_1'}\A_{a_1'a_2}^{x_1'x_2}\tp\proj{x_1'})(\sum_{a_2''}\A_{a_1a_2''}^{x_1x_2}\tp V_{x_1})\nonumber\\
=&\sum_{x_1'}\left((\sum_{a_2'}\A_{a_1a_2'}^{x_1x_2}\sum_{a_1'}\A_{a_1'a_2}^{x_1'x_2}\sum_{a_2''}\A_{a_1a_2''}^{x_1x_2})\tp V_{x_1}^\dagger\proj{x_1'}V_{x_1}\right)
\end{align}
The resulting correlations are 
\begin{align}
    &\bra{\psi}\tp\bra{0} \sum_{x_1'}\left((\sum_{a_2'}\A_{a_1a_2'}^{x_1x_2}\sum_{a_1'}\A_{a_1'a_2}^{x_1'x_2}\sum_{a_2''}\A_{a_1a_2''}^{x_1x_2})\tp V_{x_1}^\dagger\proj{x_1'}V_{x_1}\right)\ket{\psi}\tp\ket{0} \nonumber\\ =&\bra{\psi}(\sum_{a_2'}\A_{a_1a_2'}^{x_1x_2}\sum_{a_1'}\A_{a_1'a_2}^{x_1x_2}\sum_{a_2''}\A_{a_1a_2''}^{x_1x_2})\ket{\psi}=\bra{\psi}\A_{a_1a_2}^{x_1x_2}\ket{\psi}
\end{align}
as desired.

\section{Hierarchy for time ordered local correlations}\label{app:local}

Here we show how to modify the our hierarchy for sequential quantum correlations introduced in the main text in order to approximate the set of time ordered local correlations.
Following \cite{Gallego2014}, we say that the correlations from a Bell scenario are time ordered local if they can be described by the following model

\begin{equation} \label{eq:localseq}
P( \mathbf{a},\mathbf{b}\vert \mathbf{x},\mathbf{y} ) = \int_{\lambda} d\lambda \rho({\lambda}) p(\mathbf{a}\vert \mathbf{x}, \lambda)  p(\mathbf{b}\vert \mathbf{y} , \lambda ) \, ,
\end{equation} 
where the distribution $p(\mathbf{a}\vert \mathbf{x}, \lambda)$ satisfies the following sequential no-signaling constraint for all values of $\lambda$
\begin{align}\label{eq:nobackprob}
    \sum_{a_{k+1},\cdots,a_n} p( \textbf{a} \vert \textbf{x} , \lambda) - p(\textbf{a} \vert \textbf{x}',\lambda) =0 \quad&\forall a_1,\dots,a_k \\[-10pt] \nonumber
    &\forall \textbf{x},\textbf{x}' \text{ s.t } x_i=x_i', \\&\qquad\qquad(i\leq k) \nonumber , 
\end{align}
and similarly for $p(\mathbf{b}\vert \mathbf{y} , \lambda )$. Correlations in the above form are the only ones that can be achieve with classical means in a sequential Bell scenario.

It is well know that, by using the constraints in \eqref{eq:nobackprob}, the model (\ref{eq:localseq}) can be reduced to a sum over deterministic strategies, namely
\begin{equation}	\label{eq:localseq_det}
P( \mathbf{a},\mathbf{b}\vert \mathbf{x},\mathbf{y} )  = \sum_{\lambda} p(\lambda) D^{SEQ}(\mathbf{a}\vert \mathbf{x}, \lambda)  D^{SEQ}(\mathbf{b}\vert \mathbf{y} , \lambda ) \, ,
\end{equation}

where the deterministic probability distributions split into a product
\begin{equation}
D^{SEQ}(\mathbf{a}\vert \mathbf{x}, \lambda) = \prod_{k = 1}^n D(a_k \vert x_1,\ldots,x_k , \lambda)    
\end{equation}
and where the expression $D(a_k \vert x_1,\ldots,x_k , \lambda)$ corresponds to outputting deterministically $a_k = \lambda(x_1,\ldots,x_k)$ depending on the strategy given by $\lambda(.)$ and on all the inputs of previous boxes in the sequence (and similarly for Bob's strategy).

Determining whether a given distribution admits a decomposition in such a form is an instance of linear programming. Indeed, it implies checking if the distribution can be written as a convex combination of a finite amount of extremal points, represented by all the possible choices of deterministic strategies $D^{SEQ}(\mathbf{a}\vert \mathbf{x}, \lambda), D^{SEQ}(\mathbf{b}\vert \mathbf{y} , \lambda )$. This linear program quickly becomes computationally intractable, since the number of extremal points increases exponentially with the number of inputs. Moreover, for each additional box in the sequence, the scaling is even worse than the equivalent multipartite locality scenario, because the possible strategies for each box depend on the inputs of all the previous boxes.

That is why we are interested in relaxing the linear program with an SDP, in a similar spirit as in \cite{Baccari17}. In particular, the objective is to have a way of determining whether a distribution is sequentially local that, despite being a relaxation, works in many relevant cases and has a better scaling with the number of inputs/boxes. In the following we show how to do this by adapting our sequential hierarchy.
The first step is to find a particular realisation of sequentially local correlations in terms of a quantum measurement on a quantum state; namely we look for realisation of the kind
\begin{equation}\label{eq:quantum}
p( \mathbf{a},\mathbf{b}\vert \mathbf{x},\mathbf{y}) = \tr(\rho_{AB} \A_{\mathbf{a}}^{\mathbf{x}}\tp\B_{\mathbf{b}}^{\textbf{y}}) \, .
\end{equation} 

Now, it can be easily checked that correlations of the kind (\ref{eq:localseq_det}) can be reproduced by the following choice of state 

\begin{equation}\label{eq:state}
\rho_{AB} = \sum_{\lambda}p(\lambda) \ketbra{\lambda}{\lambda}^{\otimes 2}
\end{equation}

and measurements for Alice and Bob's side respectively

\begin{equation}\label{eq:detmeas}
\begin{split}
\A_{\mathbf{a}}^{\mathbf{x}} & = \sum_{\lambda} D^{SEQ}(\mathbf{a}\vert \mathbf{x}, \lambda) \, \ketbra{\lambda}{\lambda} \, , \\
B_{\mathbf{b}}^{\textbf{y}} & = \sum_{\lambda} D^{SEQ}(\mathbf{b}\vert \mathbf{y}, \lambda) \, \ketbra{\lambda}{\lambda}  \, .
\end{split}
\end{equation}

It is also easy to verify that measurements in the above form satisfy the constraints \eqref{orthog} and \eqref{noback}. In particular, the second property follows directly from the fact that the deterministic strategies $D^{SEQ}(\mathbf{a}\vert \mathbf{x}, \lambda)$ and $D^{SEQ}(\mathbf{b}\vert \mathbf{y}, \lambda)$ satisfy the no-signalling condition \eqref{eq:nobackprob}.
Moreover, since all measurement operators are diagonal in the $\ket{\lambda}$ basis it follows that $[\A_{\mathbf{a}}^{\mathbf{x}},\A_{\mathbf{a'}}^{\mathbf{x'}}]=0$ and $[\B_{\mathbf{b}}^{\mathbf{y}},\B_{\mathbf{b'}}^{\mathbf{y'}}]=0$. In other words, the set of time ordered local correlations can be obtained by means of locally commuting quantum sequential measurements. These commutativity conditions imply additional linear constraints on the moment matrix elements, expressed by some fixed matrices $G_i^{LOC}$. We can thus define the following hierarchy\\

\noindent\textbf{Hierarchy for sequential local correlations  (level k)}
\begin{align}\label{NPAlocseq}
    \text{Find }\Gamma_k \quad \text{ such that} \quad 
    &\quad\;\Gamma_k \succcurlyeq 0,\; \Gamma_k^\dagger=\Gamma_k, \; \Gamma_k^{(0,0)}=1, \\
    &\quad\tr[\Gamma_k G_i]=0 \; \forall i, \nonumber\\
    &\quad\tr[\Gamma_k G^{SEQ}_i]=0 \; \forall i, \nonumber\\
    &\quad\tr[\Gamma_k G^{LOC}_i]=0 \; \forall i, \nonumber\\
    &\quad\tr[\Gamma_k F_{i}]=P_i \; \forall i. \nonumber
\end{align}
We call $\mathcal{L}_{\text{SEQ}}^k$ the set defined at level $k$ of this hierarchy. By construction, each $\mathcal{L}_{\text{SEQ}}^k$ defines an outer approximation of the set of time ordered local correlations. The computational advantage gained by replacing a linear programming characterisation of the exact set with an SDP relaxation is clear: at each fixed level $k$, the number of variables involved in the moment matrix $\Gamma_k$ scales polynomially with the number of input choices for $x_1,\ldots,x_n$ and $y_1,\ldots,y_n$, contrarily to the exponential scaling of the linear programming. This may allow one to probe scenarios which would otherwise be practically impossible using linear programming methods. 

\section{Using the hierarchy to upper bound guessing probabilities}\label{app:random}
We first review how the standard NPA hierarchy can be used to provide upper bounds to the guessing probability \eqref{pg} (see also \cite{bancal2014,nieto2014,nieto2018}). Define the subnormalised distributions (both normalised to $p_{\text{\tiny{E}}}(e)$)
\begin{align}
    &\tilde{p}^e_{\text{\tiny{AB}}}(a,b\vert x,y)=p_{\text{\tiny{ABE}}}(a,b,e\vert x,y)=p_{\text{\tiny{E}}}(e)p_{\text{\tiny{AB}}}(a,b\vert x,y,e), \nonumber\\
    &\tilde{p}^e_{\text{\tiny{B}}}(b\vert y)=p_{\text{\tiny{BE}}}(b,e\vert y)=p_{\text{\tiny{E}}}(e)p_{\text{\tiny{B}}}(b\vert y,e)=\sum_a \tilde{p}^e_{\text{\tiny{AB}}}(a,b\vert x,y). \nonumber
\end{align}
Define $\mathcal{Q}^{p}$ to be the set of quantum correlations, subnormalised to $p$, that is, $P(a,b\vert x,y)\in \mathcal{Q}^{p}$ if $P(a,b\vert x,y)=p\,P'(a,b\vert x,y)$ for some $P'\in\mathcal{Q}$. Thus $\tilde{p}^e_{\text{\tiny{AB}}}\in\mathcal{Q}^{p(e)}$ and with this notation \eqref{pg} reads
\begin{align}\label{pg2}
    G(y^*)=\max_{\tilde{p}^e_{\text{\tiny{AB}}}}\;\sum_{e=1}^{\vert{b}\vert} \tilde{p}^e_{\text{\tiny{B}}}(e\vert y^*), \qquad \qquad
    &\sum_e \tilde{p}^e_{\text{\tiny{AB}}}(a,b\vert x,y)=P_{\text{obs}}(a,b\vert x,y) \\
    &\tilde{p}^e_{\text{\tiny{AB}}}(a,b\vert x,y)\in\mathcal{Q}^{p(e)}\;\forall e  . \nonumber
\end{align}
Define the set $\mathcal{Q}_k^{p(e)}$ in the same way as $\mathcal{Q}_k$ but changing the normalisation condition $\Gamma_k^{(0,0)}=1$ to $\Gamma_k^{(0,0)}=p(e)$ in \eqref{NPA}. Thus $\mathcal{Q}_k^{p(e)}\supseteq \mathcal{Q}^{p(e)}$. The problem \eqref{pg2} can therefore be upper bounded by relaxing the condition $\tilde{p}^e_{\text{\tiny{AB}}}(a,b\vert x,y)\in\mathcal{Q}^{p(e)}$ to $\tilde{p}^e_{\text{\tiny{AB}}}(a,b\vert x,y)\in\mathcal{Q}_k^{p(e)}$. Practically, this means that one has to consider a set of $\vert b \vert$ subnormalised moment matrices in the optimisation. 

An analogous procedure can be followed in the sequential scenario by defining the set $\mathcal{Q}^{p}_{\text{SEQ}}$, the set of subnormalised sequential correlations, and  corresponding relaxations $\mathcal{Q}^{p}_{\text{SEQ},k}$. Following the same logic as above, one arrives at the guessing probability 
\begin{align}\label{pg3}
    G(\mathbf{y}^*)=\max_{\tilde{p}^{\mathbf{e}}_{\text{\tiny{AB}}}}\;\sum_{\mathbf{e}} \tilde{p}^{\mathbf{e}}_{\text{\tiny{B}}}(\mathbf{e}\vert \mathbf{y}^*), \qquad\qquad
    &\sum_\mathbf{e} \tilde{p}^{\mathbf{e}}_{\text{\tiny{AB}}}(\mathbf{a},\mathbf{b}\vert \mathbf{x},\mathbf{y})=P_{\text{obs}}(\mathbf{a},\mathbf{b}\vert \mathbf{x},\mathbf{y}) \\
    &\tilde{p}^\mathbf{e}_{\text{\tiny{AB}}}(\mathbf{a},\mathbf{b}\vert \mathbf{x},\mathbf{y})\in\mathcal{Q}^{p(\mathbf{e})}\;\forall \mathbf{e}.  \nonumber
\end{align}
One then relaxes the condition $\tilde{p}^\mathbf{e}_{\text{\tiny{AB}}}(\mathbf{a},\mathbf{b}\vert \mathbf{x},\mathbf{y})\in\mathcal{Q}^{p}_{\text{SEQ}}$ to $\tilde{p}^\mathbf{e}_{\text{\tiny{AB}}}(\mathbf{a},\mathbf{b}\vert \mathbf{x},\mathbf{y})\in\mathcal{Q}^{p}_{\text{SEQ},k}$, thus needing as many moment matrices as the total alphabet size of $\mathbf{b}$

\end{document}